\documentstyle[nolta99]{article}
\title{
Generalized One-Dimensional Point Interaction \\
in Relativistic and Non-relativistic Quantum Mechanics
}
\author{
Takaomi Shigehara$^{\dag}$, 
Hiroshi Mizoguchi$^{\dag}$, 
Taketoshi Mishima$^{\dag}$ and 
Taksu Cheon$^{\ddag}$ 
}
\affiliation{%
\dag 
Department of Information and Computer Sciences, Saitama University, \\
Shimo-Okubo, Urawa, Saitama 338-8570, Japan \\
{\tt sigehara@ics.saitama-u.ac.jp} \\
\ddag 
Laboratory of Physics, Kochi University of Technology, \\
Tosa Yamada, Kochi 782-8502, Japan \\
{\tt cheon@mech.kochi-tech.ac.jp}
}

\begin{document}
\maketitle
\abstract
We first give the solution for the local approximation of 
a four parameter family of generalized one-dimensional 
point interactions within the framework of non-relativistic 
model with three neighboring $\delta$ functions. 
We also discuss the problem within relativistic (Dirac) 
framework and give the solution for a three parameter family. 
It gives a physical interpretation for so-called 
$\varepsilon$ potential. 
It will be also shown that the scattering properties 
at high energy substantially differ between 
non-relativistic and relativistic cases. 
\endabstract

\section{Introduction}

Since the appearance of Kronig-Penney model in solid state physics, 
the point interaction has attracted much attention 
in various fields of quantum mechanics which cover 
field theory, quantum chaos and mathematical physics. 
In one dimension, the current conservation admits 
the boundary condition at the interaction  
characterized by four parameters \cite{GK85}. 
However, it is not straightforward to approximate 
generic cases except the well-known $\delta$ potential by using 
short-range local potential model. 
To rectify the situation has been a longstanding 
problem in mathematical physics \cite{SE86a,CA93}. 
In  \cite{CS98a,SM98a},  we have introduced three neighboring 
$\delta$'s model and settled the problem with the assumption that 
the system conserves time reversal symmetry. 
In this paper, we remove this assumption and 
extend the previous results to the four parameter family.  
The second purpose is to discuss the same problem 
within a relativistic framework. 
We show that the Dirac equation with short-range 
scalar and vector potentials naturally reproduces 
a three parameter family of point interactions 
without introducing renormalization of coupling strengths,  
which is required in the non-relativistic approach. 
The relativistic approach also serves to give a unified 
view to $\delta$ and $\varepsilon$ potentials. Here 
the $\varepsilon$ potential, which has been historically 
ill-called $\delta'$, gives rise to the boundary condition 
such that the wave function has continuous first derivative 
on the right and left, but it has a jump proportional to 
the first derivative \cite{GHM80}. 
We also examine the scattering properties of the generalized 
point interactions both within the non-relativistic and 
relativistic frameworks.

The paper is organized as follows. 
We discuss non-relativistic and relativistic cases 
in Sect.II and III separately. In each, 
the boundary condition required at the interaction 
is first clarified based on the current conservation. 
We then calculate the propagator in the transfer matrix 
formalism, based on which 
we deduce a suitable approximation of 
a family of point interactions by short-range local 
potentials. 
The scattering properties are discussed at the end of 
each section. 
We summarize the present work in Sect.IV.

\section{Non-Relativistic Approach}

We first derive a general boundary condition around 
the one-dimensional point interaction from the viewpoint 
of the current conservation. 
Schr\"{o}dinger equtaion with potential $S$ reads 
\begin{eqnarray}
\label{s1}
-\frac{1}{2m}\varphi_S''(x)+S(x)\varphi_S(x)=E\varphi_S(x),
\nonumber 
\end{eqnarray}
where $m$ and $E$ are the mass and energy, 
respectively.  
The current can be written as 
$j_S={\bf \Psi}_S^{\dagger}\sigma_2 {\bf \Psi}_S$, 
where  
\begin{eqnarray}
\label{s2}
 {\bf \Psi}_S(x) =
\left( \! 
\begin{array}{c}
\varphi_S(x) \\
\frac{1}{2m}\varphi_S'(x) \\
\end{array}
\! \right), 
\nonumber 
\end{eqnarray}
and $\sigma_2$ is 
the second component of Pauli matrices;  
\begin{eqnarray}
\label{s3}
\sigma_1 \!\! = \!\! \left( \begin{array}{cc}
   0  &   1 \\
   1  &   0
\end{array} \right), \hspace{1ex}
\sigma_2 \!\! = \!\! \left( \begin{array}{cc}
   0  &  -i \\
   i  &   0
\end{array} \right), \hspace{1ex}
\sigma_3 \!\! = \!\! \left( \begin{array}{cc}
   1  &   0 \\
   0  &  -1
\end{array} \right).
\nonumber 
\end{eqnarray}
Let's put a point interaction at the origin of $x$-axis. 
The current should be conserved between both sides of the interaction.  
This leads to the connection condition characterized by 
\begin{eqnarray}
\label{s4}
{\bf \Psi}_S(+0)={\cal V}{\bf \Psi}_S(-0),    
\end{eqnarray}
where ${\cal V}$ takes a form \cite{GK85} 
\begin{eqnarray}
\label{s5}
{\cal V}=e^{i\theta}{\cal U}, 
\hspace{2ex}
{\cal U} = \left(
\begin{array}{cc}
\alpha & \beta \\
\gamma & \delta 
\end{array}
\right) \in SL(2,{\bf R}),     
\hspace{2ex}
\theta \in {\bf R}. 
\end{eqnarray}
If and only if Eq.(\ref{s5}) holds, 
we have ${\cal V}^{\dagger}\sigma_2{\cal V}=\sigma_2$, 
namely the current is conserved on both sides of the interaction.  
For example, the connection matrix  
\begin{eqnarray}
\label{s6}
{\cal V}_{\delta}(v) = 
\left(
\begin{array}{cc}
1 & 0 \\
v & 1 
\end{array}
\right)
\end{eqnarray}
corresponds to the $\delta$ potential of strength $v$, 
while its transpose represents the $\varepsilon$ potential of 
strength $v$.

In order to take into account the magnetic effect, 
we calculate the propagator at the presence of 
a homogeneous vector potential 
(spatial component in a relativistic sense), 
which induces the time reversal symmetry breaking into the system. 
By using standard minimal coupling, 
Schr\"{o}dinger equation is written as 
\begin{eqnarray}
\label{s7}
- \varphi_S''(x) + 
2iA \varphi_S'(x) + A^2 \varphi_S(x) = 
k^2 \varphi_S(x), 
\end{eqnarray}
where $k$ is the wave number ($k=\sqrt{2mE}$) and 
the vector potential $A$ is constant.  
In the transfer matrix formulation, 
Eq.(\ref{s7}) is rewritten by the first-order coupled equation  
\begin{eqnarray}
\label{s8}
{\bf \Psi}_S'(x) = {\cal H}_S {\bf \Psi}_S(x), 
\hspace{1ex}
{\cal H}_S = 
\left(
\begin{array}{cc}
          0           & 2m \\
\frac{-k^2 + A^2}{2m} & 2iA
\end{array}
\right). 
\end{eqnarray}
The solution of Eq.(\ref{s8}) is given by 
\begin{eqnarray}
\label{s9}
{\bf \Psi}_S(x) = {\cal G}_S(x-x_0) {\bf \Psi}_S(x_0) 
\nonumber 
\end{eqnarray}
with the exponential function of the matrix ${\cal H}_S x$; 
\begin{eqnarray}
\label{s10}
\hspace*{-3ex}
{\cal G}_S(x) 
& = \!& e^{iAx} \left\{
\cos ( kx ) 
\left(
\begin{array}{cc}
 1 & 0 \\
 0 & 1
\end{array}
\right) 
\right. \nonumber \\ 
\hspace*{-3ex}
&  & +  \left. 
\frac{\sin ( kx )}{k}
\left(
\begin{array}{cc}
       -iA            & 2m \\
\frac{-k^2 + A^2}{2m} & iA
\end{array}
\right) 
\right\}. 
\end{eqnarray}
In a particular case of $A=0$, 
Eq.(\ref{s10}) is reduced to the free propagator 
\begin{eqnarray}
\label{s11}
{\cal G}_S^{(0)}(x) = 
\left(
\begin{array}{cc}
\cos ( kx ) & \frac{2m}{k}\sin ( kx )\\
-\frac{k}{2m}\sin ( kx ) & \cos ( kx ) 
\end{array}
\right). 
\nonumber 
\end{eqnarray}
We can see that 
${\cal G}_S^{(0)}$ 
has eigenvalues $e^{\pm ikx}$ with the associated eigenfunction 
\begin{eqnarray}
\label{s12}
{\bf u}_S^{(\pm)}=\frac{1}{\sqrt{2}}
\left( \begin{array}{c}
1 \\ \pm i\frac{k}{2m}
\end{array} \right). 
\end{eqnarray}
Also its complex conjugate 
${\cal G}_S^{(0) \dagger}$ has 
eigenvalues $e^{\mp ikx}$ with the associated eigenfunction 
\begin{eqnarray}
\label{s13}
{\bf v}_S^{(\pm)}=\frac{1}{\sqrt{2}}
\left( \begin{array}{c}
1 \\ \pm i\frac{2m}{k}
\end{array} \right).   
\end{eqnarray}
The eigenfunctions satisfy the bi-orthogonal relations; 
\begin{eqnarray}
\label{s14}
{\bf v}_S^{(\pm) \dagger}{\bf u}_S^{(\pm)}=1, \hspace{3ex}
{\bf v}_S^{(\mp) \dagger}{\bf u}_S^{(\pm)}=0. 
\end{eqnarray}

In order to realize the connection condition (\ref{s4}) 
in the small-size limit 
of a finite-range local potential, 
we consider three nearby $\delta$'s 
located with equal distance $a$ \cite{CS98a,SM98a}. 
They are put at $x=\pm a$ and $0$ and their 
strengths are denoted by $v_{\pm}$ and $v_{0}$, respectively. 
We also add a constant vector potential $A$ 
between the two side $\delta$'s.  
The connection of ${\bf \Psi}_S$ 
between $x=-a-0$ and $x=a+0$ is given by 
\begin{eqnarray}
\label{s15}
{\bf \Psi}_S(a+0) = {\cal V}_S{\bf \Psi}_S(-a-0), 
\nonumber 
\end{eqnarray}
where
\begin{eqnarray}
\label{s16}
{\cal V}_S = 
{\cal V}_{\delta}(v_+ \!\! - i\frac{A}{2m})
{\cal G}_S(a) 
{\cal V}_{\delta}(v_0) 
{\cal G}_S(a) 
{\cal V}_{\delta}(v_- \!\! + i\frac{A}{2m}).   
\nonumber 
\end{eqnarray}
The imaginary strength of the side $\delta$'s 
indicates sudden change of the vector potential at $x=\pm a$.  
Using Eqs.(\ref{s6}) and (\ref{s10}), we reach
\begin{eqnarray}
\label{s17}
{\cal V}_S = e^{2iAa} {\cal U}_S, 
\end{eqnarray}
where ${\cal U}_S \in SL(2,{\bf R})$ has components
\begin{eqnarray}
\label{s18}
\left[{\cal U}_S\right]_{11} & = & 
\cos ( 2ka ) + \frac{m\sin ( 2ka )}{k} v_0 
+ \left[{\cal U}_S\right]_{12} v_-, \nonumber \\ 
\left[{\cal U}_S\right]_{12} & = & \frac{2m\sin ( 2ka )}{k} + 
\frac{ 4m^2\sin^2 ( ka )}{k^2} v_0, \nonumber \\
\left[{\cal U}_S\right]_{21} & = & 
\cos^2 ( ka ) \cdot (v_+ + v_0 + v_-) \nonumber \\
& & - \sin^2 ( ka ) \cdot (v_+ + v_-) \nonumber \\
& & + \frac{m\sin ( 2ka )}{k} 
\left\{ -\frac{k^2}{2m^2} + v_0 ( v_+ + v_-) \right\} \nonumber \\
& & + \left[{\cal U}_S\right]_{12} v_+ v_-, \nonumber \\
\left[{\cal U}_S\right]_{22} & = & 
\cos ( 2ka ) + \frac{m\sin ( 2ka )}{k} v_0 
+ \left[{\cal U}_S\right]_{12} v_+. \nonumber  
\nonumber 
\end{eqnarray}
The effect of the magnetic field appears only 
in the phase factor in Eq.(\ref{s17}).  
By examining the behavior of ${\cal U_S}$ for small $a$, 
we can show \cite{SM98a} that 
the matrix ${\cal V}_S$ converges 
to the general connection ${\cal V}$,   
if the strengths of the three $\delta$'s are renormalized 
according to the distance $a$ as follows;     
For $\beta\neq 0$,  
\begin{eqnarray}
\label{s19}
\left\{ \begin{array}{lll}
v_+ & = & \displaystyle -\frac{1}{2ma} + \frac{\delta+1}{\beta}, \\[2.0ex]
v_0 & = & \displaystyle \frac{\beta}{4m^2 a^2},            \\[2.0ex]
v_- & = & \displaystyle -\frac{1}{2ma} + \frac{\alpha+1}{\beta}, 
\end{array} \right. 
\end{eqnarray}
whereas for $\beta=0$ (including $\delta$ potential), 
\begin{eqnarray}
\label{s20}
\left\{ \begin{array}{lll}
v_+ & = & \displaystyle \frac{\delta-1}{4ma}, \\[2.0ex] 
v_0 & = & \displaystyle \frac{4\gamma}{\alpha+\delta+2}, \\[2.0ex]
v_- & = & \displaystyle \frac{\alpha-1}{4ma}.
\end{array} \right. 
\end{eqnarray}
In both cases, we vary the strength of the vector potential 
as $A = \frac{\theta}{2a}$. 
Eqs.(\ref{s19}) and (\ref{s20}) show that except 
the $\delta$ potential ($\alpha=\delta=1$, $\beta=0$), 
we need to renormalize the strengths of the three $\delta$'s  
to realize the connection condition. 
Setting $\alpha=\delta=1$ in Eq.(\ref{s19}), 
we obtain the $\varepsilon$ potential of strength $\beta$, 
in which case the strengths diverge in the small $a$ limit 
as in generic cases.

In the transfer matrix formulation, 
the bi-orthogonal eigenvectors (\ref{s12}) and (\ref{s13}) 
serve to examine the scattering properties. 
Since $e^{\pm ikx}{\bf u}_S^{(\pm)}$ 
are the solutions of the equation (\ref{s8}) 
for the free space ($A=0$), 
the wave function is written as 
\begin{eqnarray}
\label{s21}
{\bf \Psi}_S(x)= 
\left\{ \begin{array}{ll}
e^{ikx}{\bf u}_S^{(+)} + R_S e^{-ikx}{\bf u}_S^{(-)}, & (x<0), \\
T_S e^{ikx}{\bf u}_S^{(+)}, & (x>0), \\
\end{array}\right.
\end{eqnarray}
where $T_S$ and $R_S$ are transmission and reflection coefficients, 
respectively. 
We also assume that the incident wave comes from minus infinity 
in Eq.(\ref{s21}).   
From the connection condition (\ref{s4}), we obtain 
\begin{eqnarray}
\label{s22}
T_S {\bf u}_S^{(+)} = {\cal V} ({\bf u}_S^{(+)} + R_S {\bf u}_S^{(-)}). 
\nonumber 
\end{eqnarray}
Multiplying ${\bf v}_S^{(+) \dagger}{\cal V}^{-1}$ from the left and 
using the bi-orthogonal relations (\ref{s14}), 
we can estimate the transition probability as 
\begin{eqnarray}
\label{s23}
\hspace*{-3ex}
|T_S|^2 & 
\hspace{-1ex}
= &  
\hspace{-1ex}
\left| {\bf v}_S^{(+) \dagger} 
{\cal V}^{-1}{\bf u}_S^{(+)} \right|^{-2} \nonumber \\
& 
\hspace{-1ex}
= & 
\hspace{-1ex}
4\left[ \alpha^2+\delta^2+2 +
\beta^2 \frac{k^2}{4m^2} + \gamma^2 \frac{4m^2}{k^2} 
\right]^{-1}. 
\end{eqnarray}
The reflection probability is given by 
$|R_S|^2 = 1- |T_S|^2$. 
From Eq.(\ref{s23}), we can recognize in generic cases, 
$|T_S|^2 \longrightarrow 0$ as $k\longrightarrow 0$, 
i.e. perfect reflection in the low energy limit.  
The exception is the $\varepsilon$ potential; $\gamma=0$ 
($\alpha=\delta=1$). In this case, we have 
$|T_S|^2 \longrightarrow 1$ as $k\longrightarrow 0$, 
namely perfect transmission. 
Eq.(\ref{s23}) also shows $|T_S|^2 \longrightarrow 0$ as 
$k\longrightarrow +\infty$ in generic cases, 
namely perfect reflection at high energy.  
The exception is the $\delta$ potential, $\beta=0$ 
($\alpha=\delta=1$), in which case we have 
$|T_S|^2 \longrightarrow 1$ as $k\longrightarrow +\infty$.  
Thus perfect transmission is realized.

\section{Relativistic Approach}

We start with one-dimensional Dirac equation. 
In one dimension, Dirac equation can be written by using 
two component spinors. 
This reflects the fact that 
the spin degrees of freedom are redundant in one dimension. 
One possible representation of time-independent Dirac 
equation is \cite{GS87}
\begin{eqnarray}
\label{v1}
-i \sigma_2 {\bf \Psi}_D' + 
\sigma_3 (m+S(x)){\bf \Psi}_D = (E-V(x)){\bf \Psi}_D, 
\nonumber 
\end{eqnarray}
where scalar potential $S$ and the time component of vector potential 
$V$ are taken into account.  
The facts $\sigma_2^2=\sigma_3^2=I_2$, 
$\sigma_2\sigma_3+\sigma_3\sigma_2=O_2$ 
ensure the relation $E^2=m^2+k^2$ for the free spinor 
(in case of $S=V=0$). 
In this representation, the current is written as 
$j_D={\bf \Psi}_D^{\dagger}\sigma_2{\bf \Psi}_D$. 
Thus, as in the non-relativistic case,   
the current conservation requires 
the boundary condition at the interaction such that 
\begin{eqnarray}
\label{v2}
{\bf \Psi}_D(+0)={\cal V}{\bf \Psi}_D(-0)    
\end{eqnarray}
with ${\cal V}$ in Eq.(\ref{s5}).

To derive the relativistic propagator in homogeneous potentials,  
we assume $S$ and $V$ as everywhere constant. 
As in the non-relativistic case, 
we also introduce the spatial component of vector potential 
with constant strength $A$, 
which breaks the time reversal symmetry of the system. 
The Dirac equation with the three constant potentials reads 
\begin{eqnarray}
\label{v3}
\sigma_2 (-i\frac{d}{dx}-A){\bf \Psi}_D +\sigma_3 (m+S){\bf \Psi}_D 
= (E-V){\bf \Psi}_D.  
\end{eqnarray}
By multiplying $i\sigma_2$ from the left, 
Eq.(\ref{v3}) is rewritten as 
\begin{eqnarray}
\label{v4}
{\bf \Psi}_D'(x) = {\cal H}_D 
{\bf \Psi}_D(x), 
\hspace{3ex}
{\cal H}_D = 
\left(
\begin{array}{cc}
   iA        & \tilde{k}_+ \\ 
-\tilde{k}_- & iA
\end{array}
\right), 
\end{eqnarray}
where $\tilde{k}_+ = m+E+S-V$ and 
$\tilde{k}_- = E-m-S-V$.  
In the free space ($S=V=A=0$), we have 
\begin{eqnarray}
\label{v5}
{\cal H}_D^{(0)} = 
\left(
\begin{array}{cc}
  0  &  k_+ \\ 
-k_- &  0 
\end{array}
\right),  
\end{eqnarray}
where $k_+=m+E$, $k_-=E-m$. 
A Remarkable notice is that at low energy $E\simeq m$,  
since $k_+ \simeq 2m$ and $k_- \simeq \frac{k^2}{2m}$, 
we realize ${\cal H}_D^{(0)} \simeq {\cal H}_S$ 
in Eq.(\ref{s8}) (with $A=0$);    
The relativistic problem (\ref{v3}) is formally 
equivalent to the non-relativistic one (\ref{s8}) at low energy 
in the free space. 
This indicates that the lower component of the low-energy 
free Dirac spinor is just given by the derivative 
of the upper component (within some factor), 
although the lower component is 
a dynamical variable in Dirac approach. 
This admits us to identify a point interaction in Dirac formalism 
with the corresponding non-relativistic one which satisfies 
the same boundary condition, at least at low energy.

The solution of Eq.(\ref{v4}) is written as 
\begin{eqnarray}
\label{v6}
{\bf \Psi}_D(x) = {\cal G}_D (x-x_0) {\bf \Psi}_D(x_0),   
\nonumber 
\end{eqnarray}
where the propagator ${\cal G}_D$ is defined by the exponential function of 
${\cal H}_D x$ and the result is given by 
\begin{eqnarray}
\label{v7}
{\cal G}_D(x) 
= e^{iAx} \left( \begin{array}{cc}
\cos ( \tilde{k}x ) & \frac{\tilde{k}_+}{\tilde{k}}\sin(\tilde{k}x)\\
-\frac{\tilde{k}_-}{\tilde{k}}\sin(\tilde{k}x) & \cos(\tilde{k}x) 
\end{array} \right) 
\end{eqnarray}
for $\tilde{k}_+\tilde{k}_->0$ and 
\begin{eqnarray}
\label{v8}
{\cal G}_D(x) 
= e^{iAx} \left( \begin{array}{cc}
\cosh( \tilde{k}x ) & \frac{\tilde{k}_+}{\tilde{k}}\sinh(\tilde{k}x)\\
-\frac{\tilde{k}_-}{\tilde{k}}\sinh(\tilde{k}x) & \cosh(\tilde{k}x) 
\end{array} \right) 
\end{eqnarray}
for $\tilde{k}_+\tilde{k}_-<0$ 
with $\tilde{k}=\sqrt{|\tilde{k}_+\tilde{k}_-|}$.  
Setting $S=V=A=0$ in Eq.(\ref{v7}), we obtain the free propagator 
\begin{eqnarray}
\label{v9}
{\cal G}_D^{(0)}(x) = 
\left(
\begin{array}{cc}
\cos ( kx ) & \frac{k_+}{k}\sin ( kx )\\
-\frac{k_-}{k}\sin ( kx ) & \cos ( kx ) 
\end{array}
\right),     
\nonumber 
\end{eqnarray}
where $k=\sqrt{k_+ k_-}=\sqrt{E^2-m^2}$. 
The propagator 
${\cal G}_D^{(0)}$ 
has eigenvalues $e^{\pm ikx}$ with the associated eigenfunction 
\begin{eqnarray}
\label{v10}
{\bf u}_D^{(\pm)}=\frac{1}{\sqrt{2}}
\left( \begin{array}{c}
1 \\ \pm i\frac{k_-}{k}
\end{array} \right), 
\nonumber 
\end{eqnarray}
and ${\cal G}_D^{(0) \dagger}$ has 
eigenvalues $e^{\mp ikx}$ with the associated eigenfunction 
\begin{eqnarray}
\label{v11}
{\bf v}_D^{(\pm)}=\frac{1}{\sqrt{2}}
\left( \begin{array}{c}
1 \\ \pm i\frac{k_+}{k}
\end{array} \right).     
\nonumber 
\end{eqnarray}
These eigenfunctions satisfy the bi-orthogonal relations  
as in the non-relativistic case. 
Note that $e^{\pm ikx}{\bf u}_D^{(\pm)}$ 
corresponds to the free spinor by construction.

To realize a family of generalized point interactions, 
we consider Dirac equation with a single step-like barrier potential.  
We assume that potential strengths take constants  
$S$, $V$ and $A$ in the interval $-a\leq x\leq a$ and otherwise zero. 
The connection condition is given by 
\begin{eqnarray}
\label{v12}
{\bf \Psi}_D(a+0) = {\cal V}_D{\bf \Psi}_D(-a-0), \hspace{2ex}
{\cal V}_D={\cal G}_D(2a). 
\nonumber 
\end{eqnarray}
We set $\tilde{p}_{\pm}=\pm s-v$ in the following. 
In the small-size limit with keeping 
$s=2aS$, $v=2aV$ and $\theta=2aA$ constant,  
we obtain for $\tilde{p}_+\tilde{p}_->0$ ($s^2<v^2$), 
\begin{eqnarray}
\label{v13}
\lim_{a\rightarrow 0}{\cal V}_D = 
e^{i\theta} \left( \begin{array}{cc}
\cos\tilde{p} & \frac{\tilde{p}_+}{\tilde{p}}\sin\tilde{p}\\
-\frac{\tilde{p}_-}{\tilde{p}}\sin\tilde{p} & \cos\tilde{p}
\end{array} \right), 
\end{eqnarray}
and for $\tilde{p}_+\tilde{p}_-<0$ ($s^2>v^2$),  
\begin{eqnarray}
\label{v14}
\lim_{a\rightarrow 0}{\cal V}_D = 
e^{i\theta} \left( \begin{array}{cc}
\cosh\tilde{p} & \frac{\tilde{p}_+}{\tilde{p}}\sinh\tilde{p}\\
-\frac{\tilde{p}_-}{\tilde{p}}\sinh\tilde{p} & \cosh\tilde{p}
\end{array} \right) 
\end{eqnarray}
with $\tilde{p}=\sqrt{|\tilde{p}_+\tilde{p}_-|}$. 
The present model obviously simulates a  
relativistic $\delta$ potential characterized by 
three strengths $s$, $v$ and $\theta$,  
which is sufficient to realize a three-parameter family of 
the generalized point interactions in Eqs.(\ref{v13}) and (\ref{v14}). 
This makes a clear contract to the non-relativistic case, 
where three $\delta$'s 
together with renormalization of the strengths  
are necessary to produce generic point interactions except $\delta$.

In the absence of the spatial component of vector potential 
($\theta=0$), we obtain $\delta$ connection with 
strength $-p_- =2s$ for $p_+=0$ ($s=v$), 
while $\varepsilon$ connection with 
strength $p_+ =2s$ for $p_-=0$ ($s=-v$). 
Keeping in mind the notice mentioned below Eq.(\ref{v5}), 
we realize $\delta$ and $\varepsilon$ potentials from  
a unified viewpoint within Dirac model with 
scalar and vector potentials; 
Non-relativistic $\delta$ (resp. $\varepsilon$) potential 
is nothing but sum (resp. subtraction) 
of zero-range scalar and vector potentials with common strength. 
It clarifies a relativistic origin of the $\varepsilon$ potential, 
which requires a complicated procedure for construction within the 
non-relativistic framework. 
Once establishing the $\delta$ and $\varepsilon$ potentials, 
one can realize the general boundary condition (\ref{v2}) 
by using matrix decomposition of connection matrix 
\cite{CS98a,SM98a}.

In a similar manner as in Eq.(\ref{s23}), we obtain 
the transition probability for Dirac cases; 
\begin{eqnarray}
\label{v15}
|T_D|^2= 4\left[ \alpha^2+\delta^2+2+
\beta^2 \frac{E-m}{E+m} +\gamma^2 \frac{E+m}{E-m}
\right]^{-1}. 
\nonumber 
\end{eqnarray}
At low energy, $|T_D|^2 \simeq |T_S|^2$, as expected. 
However, $|T_D|^2$ substantially differs from $|T_S|^2$ 
in the high energy limit; 
Generic cases including $\delta$ and $\varepsilon$ connections 
give rise to the non-zero transmission probability. 
In particular, the perfect transmission is attained by 
pure vector potential $s=0$ 
($\alpha=\delta=\cos v$ and $\gamma=-\beta=\sin v$).

\section{Conclusion}

We have discussed the quantum-mechanical 
generalized one-dimensional point interactions
both in the non-relativistic and relativistic (Dirac) frameworks. 
In the non-relativistic approach, 
three nearby $\delta$'s make it possible to 
construct the short-range local approximation 
for a four parameter family  
which exhausts all possible current-conserved point interactions. 
In Dirac approach, a single short-range step-like barrier 
successfully describes a three parameter family including 
$\delta$ and $\varepsilon$ potentials. 
The model gives a simple interpretation of 
$\varepsilon$ in terms of the strengths 
of scalar and vector potentials. 
The scattering properties substantially differ at high energy 
between non-relativistic and relativistic cases; 
In non-relativistic approach, generic cases lead to 
perfect reflection in the high energy limit except 
the $\delta$ potential, in which case perfect transmission 
is realized. In Dirac approach, the transmission remains 
at high energy in generic cases except pure vector potential, 
which induces perfect transmission.

\end{document}